\documentclass{ws-procs961x669}            

\usepackage{gensymb}

\input{aasmacros.sty}

\def\figsubcap#1{\par\noindent\centering\footnotesize(#1)}

\begin{document}

\title{The SVOM Mission}

\author{On behalf of the \textit{SVOM} collaboration}

\author{J-L. Atteia$^{1}$, B. Cordier$^{2}$, J. Wei$^3$}, 

\address{$^1$ IRAP, Universit\'e de Toulouse, CNRS, CNES, UPS, (Toulouse), France\\
E-mail: jean-luc.atteia@irap.omp.eu}

\address{$^2$ Lab AIM - CEA, CNRS, Universit\'e Paris-Saclay, \\
Universit\'e de Paris, 91191, Gif-sur-Yvette, France\\
E-mail: bertrand.cordier@cea.fr}

\address{$^3$ Key Laboratory of Space Astronomy and Technology, National Astronomical Observatories, Chinese Academy of Sciences, Beijing 100101, People’s Republic of China\\
E-mail: wjy@nao.cas.cn}

\begin{abstract}
The Sino-French space mission \textit{SVOM} is mainly designed to detect, localize and follow-up Gamma-Ray Bursts and other high-energy transients. 
The satellite, to be launched mid 2023, embarks two wide-field gamma-ray instruments and two narrow-field telescopes operating at X-ray and optical wavelengths. 
It is complemented by a dedicated ground segment encompassing a set of wide-field optical cameras and two 1-meter class follow-up telescopes.
In this contribution, we describe the main characteristics of the mission and discuss its scientific rationale and some original GRB studies that it will enable.\end{abstract}


\bodymatter

\section{Why a new Gamma-Ray Burst Mission?}
\label{sec:intro}
\textit{SVOM} (Space-based multi-band astronomical Variable Objects Monitor) is a Sino-French mission devoted to the study of gamma-ray bursts (GRBs) to be launched in 2023.\citep{Wei2016}
Its main goals are the detection and multi-wavelength study of high-energy transients, and the search for the electromagnetic counterparts of non-photonic transient sources detected in gravitational waves or neutrinos.
Before describing \textit{SVOM} in \Sref{sec:svom}, we briefly discuss in this section the rationale for a new GRB mission in the coming years.

\subsection{High-Energy Astrophysics}
\label{sub:hea}
The astrophysics of high-energy transients is essential for the study of several major questions of modern astrophysics.

High-energy transients offer a unique view at extreme stellar explosions with relativistic jets. High-energy observations reveal the presence of a 'central engine' able to launch relativistic jets, and as such they are essential to clarify key issues like the complex zoology of GRBs (which encompasses long and short duration GRBs, low-luminosity GRBs and ultra-long GRBs), the nature of the GRB-supernova (SN) connection or the distribution of GRB beaming angles. 
The jets themselves raise several unanswered questions, like their composition and geometry, the nature of the central engine (a magnetar or a black hole?) and the mechanism of their acceleration, the physical processes explaining their high radiative efficiency, and their role in the acceleration of Very High-Energy cosmic rays. The jets have their origin in the universal accretion / ejection mechanism that takes place around accreting compact objects. However, GRB jets being extremely relativistic with Lorentz factors of the order of one hundred, they probe a regime that is not accessible with other phenomena.

Some high-energy transients are associated with the birth of stellar mass black holes (BHs). BH astrophysics has emerged as a major topic after the discovery of dozens of binary black hole mergers by the LIGO Scientific Collaboration and Virgo Collaboration (LVC) \citep{Abbott2021}. Black hole demography and zoology, their mass \& spins distribution and their birth places are key questions which are now actively studied. In this context, GRBs are of high interest because they offer a complementary view at the birth of stellar mass BHs across the history of the universe. 

Gamma-ray bursts are so intense that some of them are detectable out to high redshift (z $\ge 5$). They illuminate regions of the young universe (at z=6 the universe is about 950 Myr old), pinpointing galaxies that would otherwise remain undetectable, allowing to measure their chemical enrichment and physical state, as well as the properties of the intergalactic medium at such high redshifts. Additionally, the history of GRB formation provides crucial insight into the formation rate of massive stars in the early universe, and pushing the redshift limit a bit further may permit the identification of GRBs from the first generation of stars (usually called population III stars) giving us a glimpse at the formation of the first stellar mass black holes. The current GRB samples probably contain a small fraction of very distant GRBs that have not been recognized in the absence of NIR spectra of their afterglows, stressing the need for fast NIR followup of GRB afterglows.  
    
Neutron stars (NSs) are prolific sources of high-energy transients, which can be emitted by very different mechanisms in accreting neutron stars, magnetars and in mergers of neutron stars. Fundamental questions like the equation of state of neutron stars, the mass gap between neutrons stars and black holes, or the radiation process of magnetars are active fields of research which require continued observations of violent phenomena involving neutron stars.

Finally, GRBs are powerful tools to test the Lorentz invariance over very large distances and they may also become a privileged tool for cosmography if we manage to standardize them. 

\subsection{Multi-messenger Astrophysics}
\label{sub:mma}
The detection of transient gravitational wave signals from cosmic explosions has opened a new window on violent phenomena involving compact objects like neutron stars and black holes. 
The complementarity of gravitational wave (GW) astrophysics and HE astrophysics has been wonderfully illustrated by the joint detection of GW170817 and GRB~170817A \citep{Abbott2017} followed by the discovery of the kilonova AT~2017gfo. These discoveries opened the era of multi-messenger astrophysics.
This new context raises new questions for HE astrophysicists, like the geometry of the jets of short GRBs, the compared demography of BNS mergers and short GRBs, the compared demography of BBH mergers and long GRBs, the masses of BHs involved in galactic HE transients, in BBH mergers and in GRBs, the photonic emission of BNS mergers, BHNS mergers, and BBH mergers, the delay between a merger and the emergence of a GRB, or the comparison of the speed of gravitational waves with the speed of light.
This non-comprehensive list illustrates a renewed interest for transient HE observations in the multi-messenger era.

\subsection{A Rich Instrumental Panorama}
\label{sub:pano}

Additional motivations to develop new instruments for the exploration of the transient high-energy sky are the diversity of sources that emit bursts of HE photons, the fact that some of them are rare events that do not repeat, and the rapid evolution of the instrumental panorama.

The transient manifestations of black holes at high energies involve galactic and extragalactic X-ray binaries, GRBs of all types, Active Galactic Nuclei (AGNs), Relativistic Tidal Disruption Events (TDEs), and possibly BHNS mergers. 
Neutron stars, on the other hand, manifest themselves as galactic and extragalactic X-ray binaries, magnetars or soft gamma repeaters eventually associated with FRBs, and of course as BNS mergers.
The diversity of these manifestations calls for different observing strategies, in terms of energy range, sensitivity, pointing strategy, follow-up, etc., which cannot be accommodated with a single mission. 

The construction of novel missions is further justified by the fact that the coming years are expected to be a golden age for transient sky astronomy and for multi-messenger astrophysics, with a number of large facilities surveying the sky for photonic signals at all wavelengths and for non-photonic messengers.
In the electromagnetic domain, we can mention the Cherenkov Telescope Array (CTA\cite{cta2019}), the High-Altitude Water Cherenkov Observatory (HAWC\cite{Albert2020}) and the Large High Altitude Air Shower Observatory (LHAASO\cite{Bai2019}) for Very High-Energy gamma-rays. In X-rays, eROSITA \citep{Predehl2021} will establish a reliable map of the X-ray sky, allowing the fast identification of new sources. In the visible the already operating Zwicky Transient Facility (ZTF\cite{ztf2019}) and the future Vera Rubin Observatory \citep{lsst2019} will map the visible transient sky every night allowing meaningful comparisons with the transient high-energy sky. In the meantime, new spectrometers on large telescopes, like SOXS\citep{soxs2016} or NTE\footnote{https://nte.nbi.ku.dk/} will increase our capacity to measure the redshifts of extragalactic transients.
In the radio domain, the counterparts of HE transients will be observed with very good sensitivity with the precursors of SKA, like ASKAP\citep{askap2008} or MEERKAT\citep{meerkat2009}, while instruments like the Canadian Hydrogen Intensity Mapping Experiment (CHIME\cite{chime2018}) or the Commensal Real-time ASKAP Fast Transients Survey (CRAFT\cite{craft2010}) will monitor the transient radio sky for Fast Radio Bursts (FRBs).

The true revolution, however, comes from the instruments looking for non-photonic messengers from violent events, in particular the gravitational waves interferometers LIGO, VIRGO, GEO and KAGRA, which have opened the exploration of the transient gravitational wave universe\citep{Abbott2020}.
As shown with the detection of GW170817 and GRB~170817A, the joint operation of gravitational wave detectors and HE detectors brings crucial information on transient gravitational wave sources.
The field of neutrino astronomy is also evolving very quickly and the large neutrinos observatories like ICECUBE\cite{IceCube2013} or KM3NeT\cite{km3net2016} have started to constrain the physics of GRB jets\cite{Senno2016}.


After this short introduction, we now present the SVOM mission (\Sref{sec:svom}) and some of its science capabilities for the study of GRBs (\Sref{sec:svsci}). 

\section{The \textit{SVOM} Mission}
\label{sec:svom}

\begin{figure}
\includegraphics[width=5in]{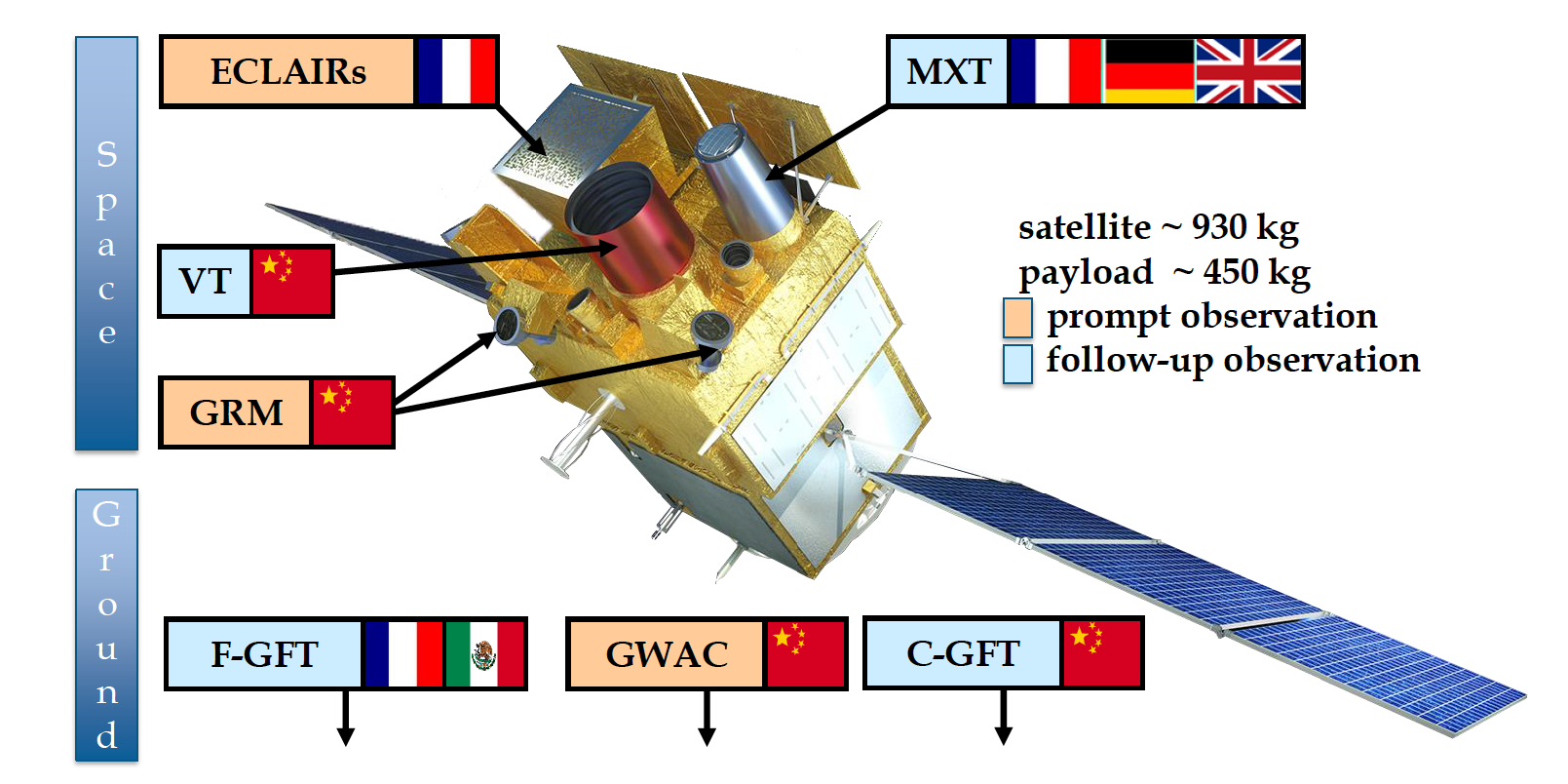}
\caption{Outline of the \textit{SVOM} mission, showing the satellite and the space and ground instruments.}
\label{fig:svom}
\end{figure}

The \textit{SVOM} mission (\Fref{fig:svom}) is the result of a bilateral collaboration between France (CNES) and China (CAS, CNSA)\cite{Gonzalez2018, Mercier2014}. It involves several research institutes from these two countries and contributions from the University of Leicester, the Max Planck Institut f\"ur Extraterrestische Physik and the Universidad Nacional Aut\'onoma de M\'exico.
 \textit{SVOM} is led by J.Y. Wei from NAOC in China and B. Cordier from CEA in France.
The mission has been designed to survey the high-energy sky and to follow-up cosmic transients at optical and X-ray wavelengths.
It encompasses a space segment, with four instruments embarked onboard a low Earth orbit satellite and a strong ground segment\cite{Chaoul2018} with two sets of wide-field optical cameras, two 1-meter class ground follow-up telescopes, and a network of $\sim45$ VHF receiving stations distributed along the footprint of the orbit.
The launch of \textit{SVOM} is scheduled mid 2023 with three years of nominal operations and a possible extension of two years.  

The main science drivers of \textit{SVOM} are high-energy transient astrophysics with a focus on gamma-ray bursts, multi-messenger astrophysics, and time domain astronomy.
The ``core program'' is dedicated to GRB detection and follow-up, aiming to improve our understanding of these phenomena.
Between GRBs, \textit{SVOM} will carry out other programs driven by the narrow-field instruments: a target of opportunity program focused on multi-messenger astrophysics and a general program focused on multi-wavelength astronomy.  
A detailed description of the \textit{SVOM} mission and its science objectives can be found in Wei et al.\cite{Wei2016} and in the website of the mission\footnote{http://www.svom.fr/}. We now briefly describe the space and ground instruments developed for \textit{SVOM}.

\subsection{Space-based Instruments}
\label{sub:space}

\subsubsection{Gamma-Ray Monitor -- GRM}
\label{ssub:grm}

\begin{figure}[ht]
\begin{center}
  \parbox{3.in}{\includegraphics[width=2.8in]{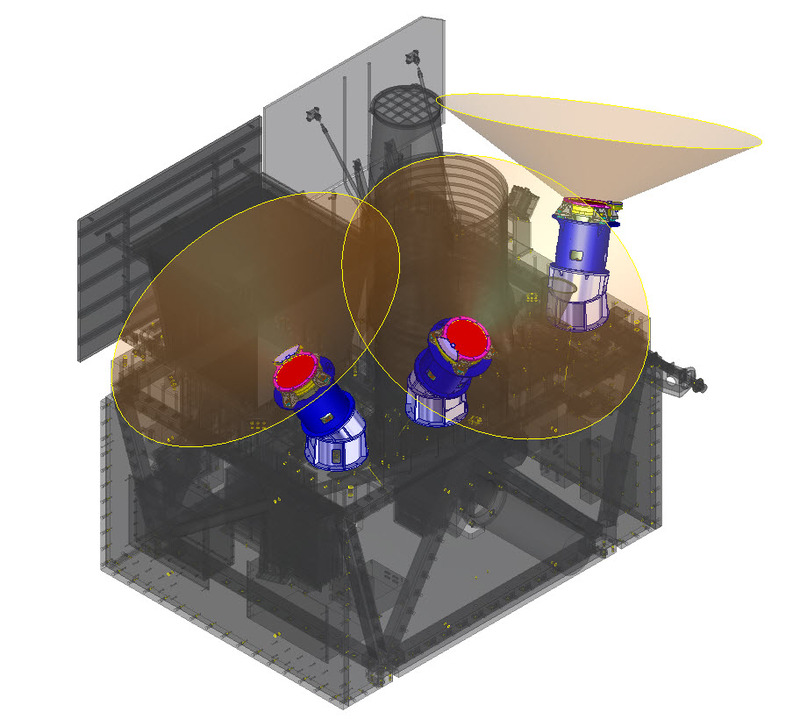}\figsubcap{a}}
  \hspace*{4pt}
  \parbox{1.8in}{\includegraphics[width=1.5in]{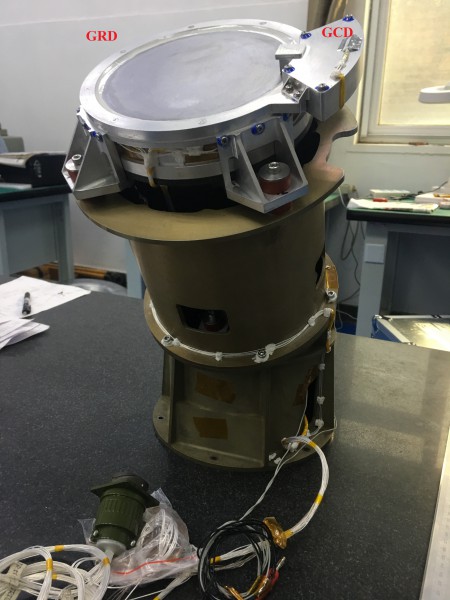}\figsubcap{b}}
  \caption{The GRM instrument. (a) Schematic drawing showing the position and orientation of the three GRD detectors on the science platform. (b) One of three GRD detection modules of the GRM.}%
  \label{fig:grm}
\end{center}
\end{figure}

The Gamma-ray monitor (GRM) is a wide-field gamma-ray spectrometer\cite{Dong2010, Zhao2013, Wen2021}. It encompasses three NaI-based detection modules (called GRDs) looking at different regions of the sky. The pointing directions of the three GRDs are 30\degree\ from the pointing axis of the satellite and they are separated by 120\degree\ in azimuth (\Fref{fig:grm}).
The main characteristics of the instrument are given in \tref{tab:grm}.
The GRM allows the detection of GRBs with a good sensitivity within a large field of view (2.6 sr), a feature which is especially important for the detection of short/hard GRBs in the era of transient gravitational wave detectors. It will also nicely complement ECLAIRs to measure jointly the GRB spectral parameters over a large energy range (4 keV - 5 MeV)\cite{Bernardini2017}.

\begin{table}
\tbl{GRM main characteristics.}
{\begin{tabular}{ll}
\toprule
Energy range (keV)  &  15 -- 5000 \\
Number of GRD modules & 3  \\
Detection area (cm$^{2}$, 1 of 3 modules) & 200  \\
Field of View (sr, all 3 modules) & 2.6  \\
Expected GRB rate (yr$^{-1}$) & 90 ($\sim 17\%$ short\cite{vonKienlin2020}) \\
Additional features & 14 detection times from 5~ms to 40~seconds in 5 energy bands. \\
& 6 mm plastic scintillator to monitor particles \& reject particle events. \\
 & Crude localization capability.  \\
\botrule
\end{tabular}}
\label{tab:grm}
\end{table}

\subsubsection{ECLAIRs}
\label{ssub:ecl}

\begin{figure}[ht]
\begin{center}
  \parbox{2.4in}{\includegraphics[width=2.3in]{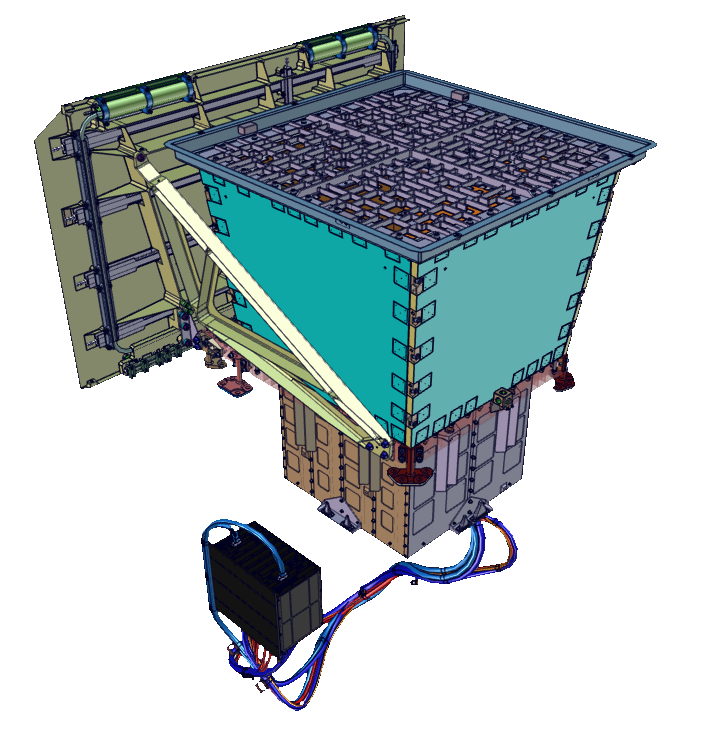}\figsubcap{a}}
  \hspace*{4pt}
  \parbox{2.4in}{\includegraphics[width=2.3in]{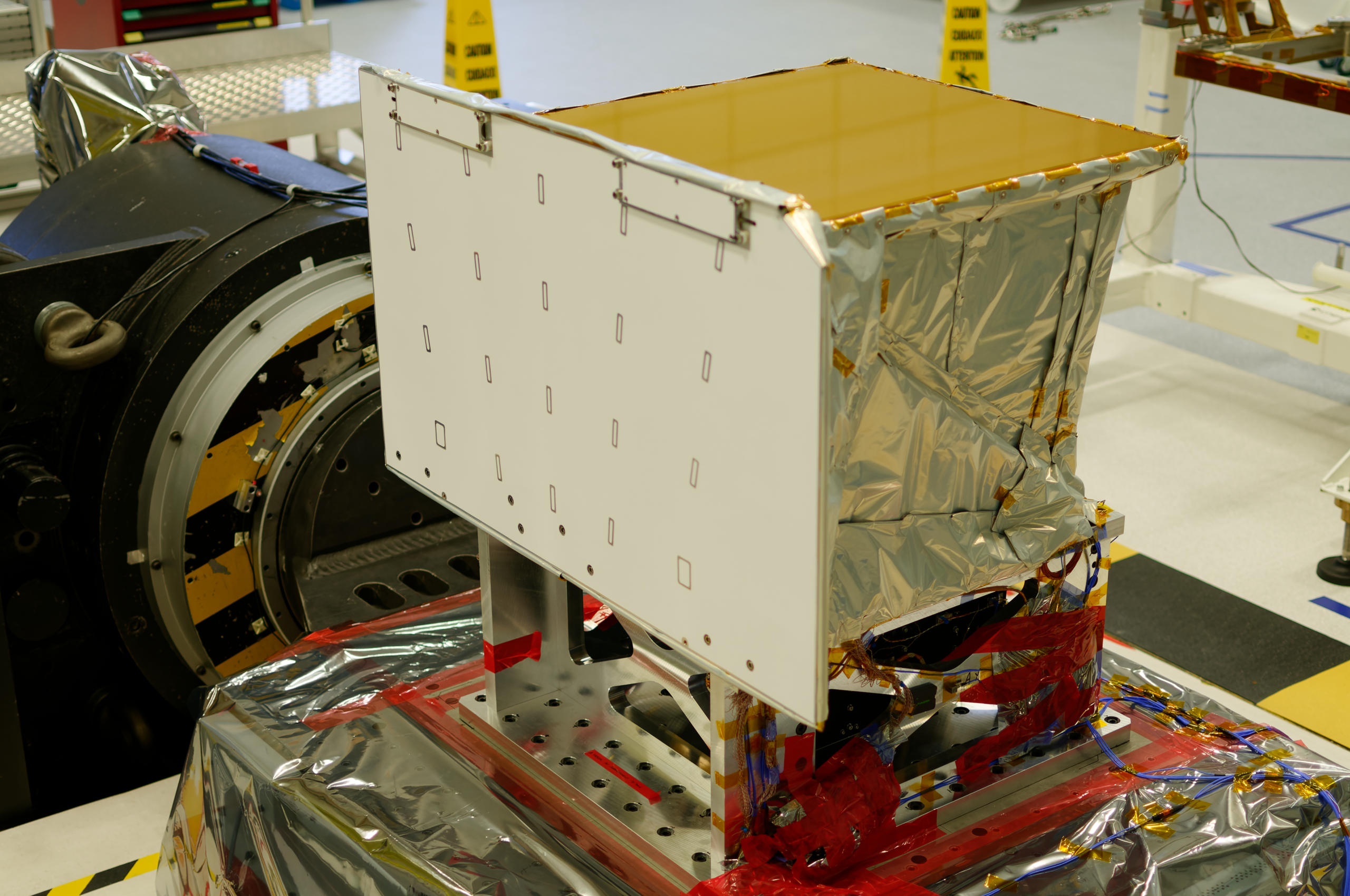}\figsubcap{b}}
  \caption{The ECLAIRs instrument. (a) Schematic drawing showing the radiator on the left, the coded mask (dark blue), the shield (light blue), the detection plane electronics (grey) and the onboard calculator (black). The detection plane itself is not visible, it is located 45~cm below the coded mask at the bottom of the shield. (b) ECLAIRs during vibration tests.}%
  \label{fig:ecl}
\end{center}
\end{figure}

ECLAIRs (\fref{fig:ecl}) is the hard X-ray imager and trigger of SVOM. The instrument is based on a coded-mask placed in front of a CdTe detection plane with 6400 4x4~mm$^{2}$ pixels\cite{Godet2014, Schanne2019}, its main characteristics are summarized in \tref{tab:ecl}.
An interesting feature of ECLAIRs is its low energy threshold at 4~keV.
Reaching this threshold was a significant technological challenge for the detectors and their readout electronics, but also for the mask, whose holes have to be transparent at these energies, and for the trigger software which has to deal with non-GRB transient sources (e.g. X-ray bursters), which are numerous at energies of a few keV. When it detects an excess in a sky image, ECLAIRs sends an alert to the satellite in order to start a slew towards the source (if the conditions are met) and to start followup with the onboard narrow-field instruments MXT \& VT.
In parallel, the alert is broadcast by the onboard VHF antennas allowing to trigger followup observations from the ground.
All the photons detected by ECLAIRs are sent to the ground, allowing to look for faint transients with improved detection methods after a typical delay of few hours.

\begin{table}
\tbl{ECLAIRs main characteristics.}
{\begin{tabular}{ll}
\toprule
Energy range (keV)  &  4.0 -- 150.0 \\
Detection area (cm$^{2}$) & 1000  \\
Number of CdTe detectors & 6400  \\
Median FWHM at 60 keV (keV) & 1.3  \\
Field of View (total, sr) & 2.0  \\
Mask open fraction (\%) & 40  \\
Localisation accuracy at detection limit (arcminute) & 12.0  \\
Expected GRB rate (yr$^{-1}$) & 65  \\
Additional features & 18 detection times from 10~ms to 20~min in 3 energy bands. \\ 
\botrule
\end{tabular}}
\label{tab:ecl}
\end{table}

\subsubsection{Microchannel X-ray Telescope -- MXT}
\label{ssub:mxt}

\begin{figure}[ht]
\begin{center}
  \parbox{1.8in}{\includegraphics[width=1.8in]{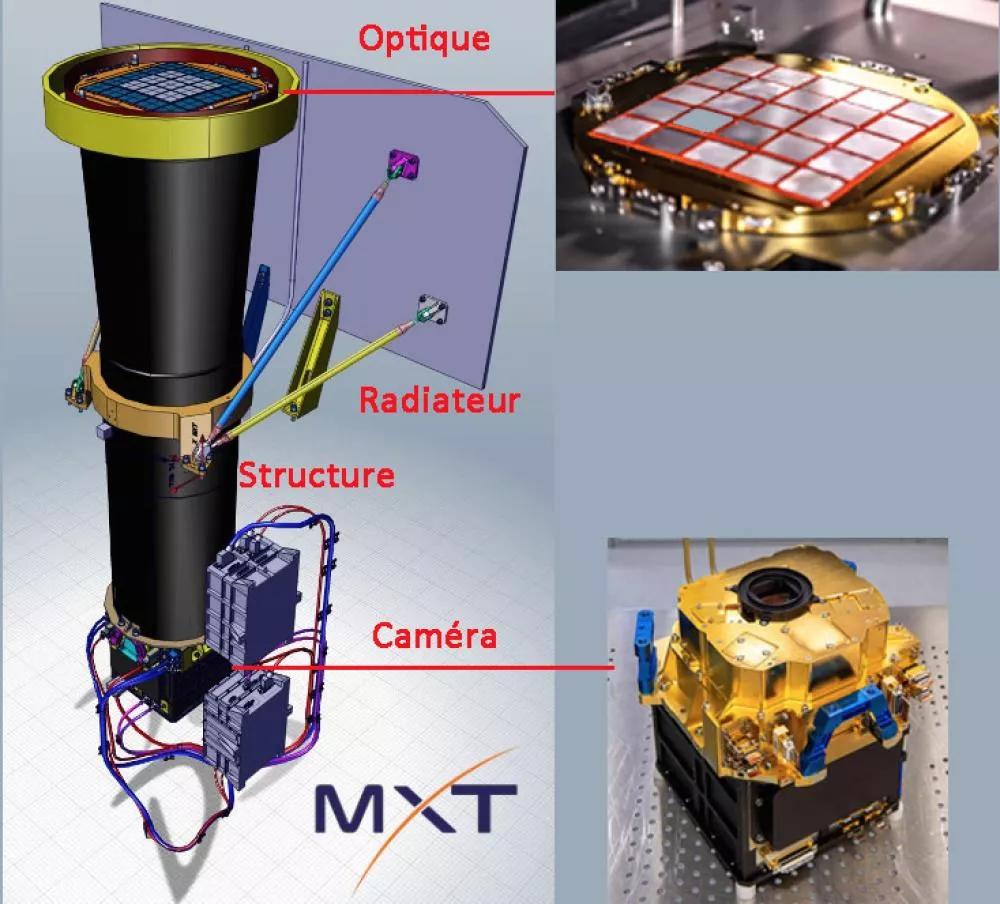}\figsubcap{a}}
  \hspace*{4pt}
  \parbox{3.in}{\includegraphics[width=2.9in]{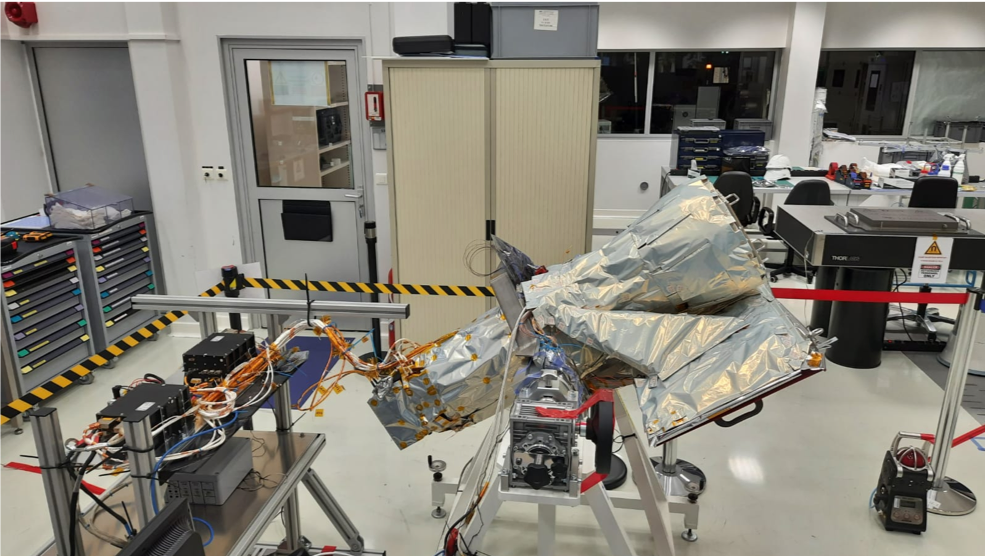}\figsubcap{b}}
  \caption{The MXT instrument. (a) Schematic drawing showing the radiator in light blue, the lobster eye optics (top inset), the shield (black), the camera (bottom inset), and the onboard calculator. (b) MXT during tests.}%
  \label{fig:mxt}
\end{center}
\end{figure}

The Microchannel X-ray Telescope (MXT, \fref{fig:mxt}) is a focusing X-ray telescope using an innovative focusing ``Lobster-Eye'' micro-pores optics and a fully-depleted frame-store silicon pnCCD camera read out by two ASICs\cite{Gotz2014, Mercier2018}. Its main characteristics are summarized in \tref{tab:mxt}.
The MXT will observe the X-ray afterglow promptly, improving ECLAIRs localization by a factor $\geq 10^{3}$.

\begin{table}
\tbl{MXT main characteristics.}
{\begin{tabular}{ll}
\toprule
Energy range (keV)  &  0.2 -- 10.0 \\
Focal length (cm) & 100  \\
Field of View (arcminute) &  64 x 64 \\
Effective area (cm$^{2}$ at 1~keV) & 27  \\
Energy resolution at 1.5 keV (eV) & 80  \\
Localisation accuracy (arcsecond) & 13.0  \\
(for 50\% of GRBs, within 5 minutes of the trigger) &  \\
Additional features & Micro-pores ``Lobster Eye'' optics ~ with square 40 $\mu$m pores. \\ 
& Silicon pnCCD based camera. \\
\botrule
\end{tabular}}
\label{tab:mxt}
\end{table}

\subsubsection{Visible Telescope -- VT}
\label{ssub:vt}

\begin{figure}[ht]
\begin{center}
  \parbox{2.0in}{\includegraphics[width=1.7in]{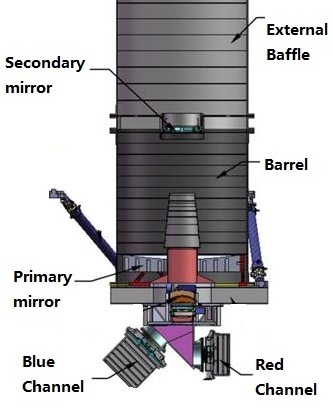}\figsubcap{a}}
  \hspace*{4pt}
  \parbox{2.8in}{\includegraphics[width=2.6in]{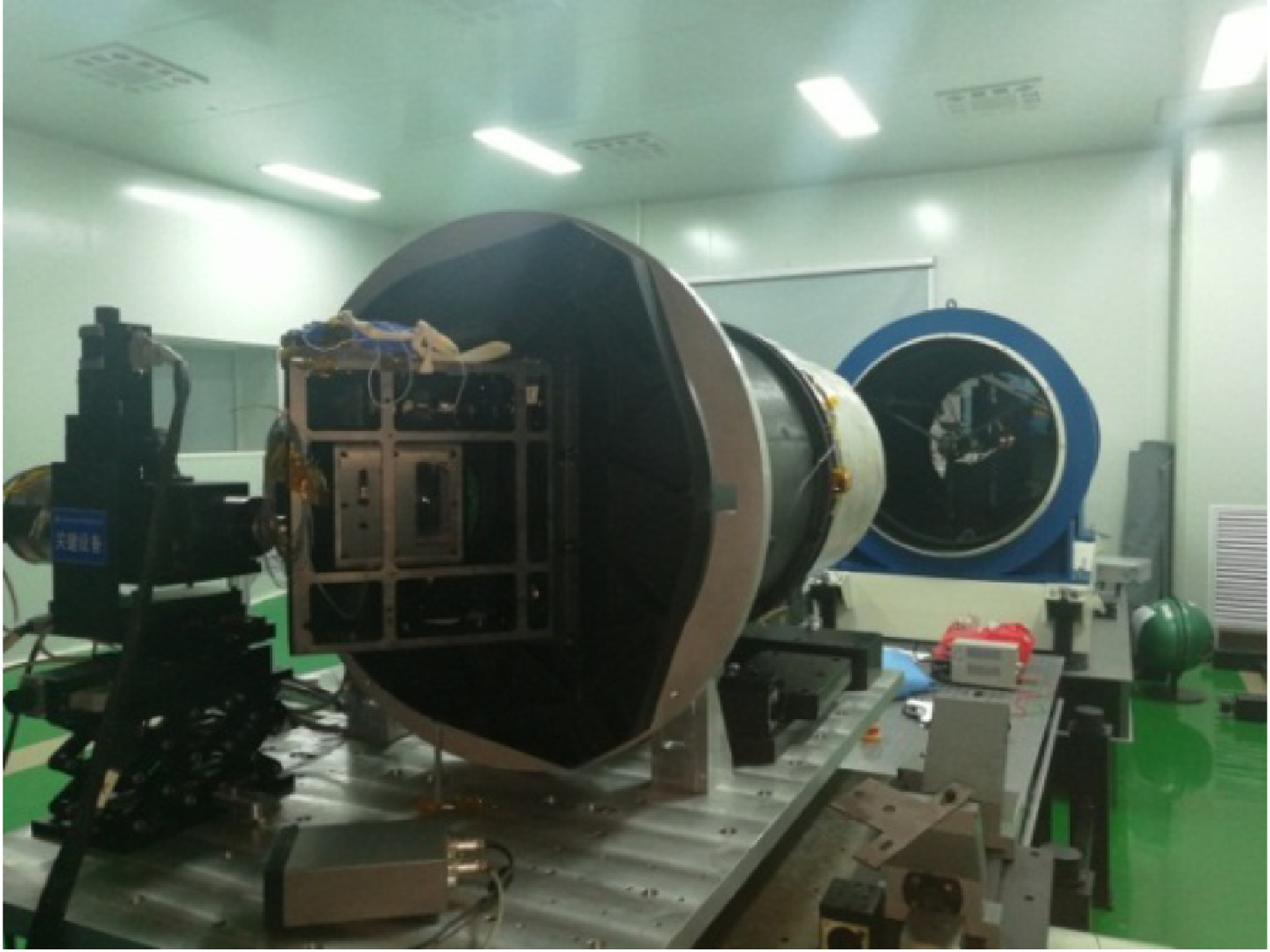}\figsubcap{b}}
  \caption{The VT instrument. (a) Schematic drawing showing the various components except the onboard calculator. (b) VT during tests.}%
  \label{fig:vt}
\end{center}
\end{figure}

The Visible Telescope (VT, \fref{fig:vt}) is a Ritchey-Chretien optical telescope with two channels, able to detect GRBs up to redshift z $\sim6.5$\cite{Wu2012, Fan2020}. It will observe SVOM GRBs quickly (in less than 5 minutes), providing an arcsecond localization of the afterglow or a deep limit ($M_{v} = 22.5$) very quickly after the burst. 

\begin{table}
\tbl{VT main characteristics.}
{\begin{tabular}{ll}
\toprule
Diameter (mm)  &  400 \\
Focal length (mm) & 3600  \\
Two channels & blue (400-650 nm) and red (650-1000 nm)  \\
Field of View (arcminute) & 26 x 26  \\
Localisation accuracy (arcsecond) & $\le 1.0$  \\
Sensitivity (300~s integration) & $M_{v} = 22.5$  \\
Additional features & Ritchey-Chretien telescope. \\ 
& Two 2k x 2k CCD detectors. \\
& Will detect 80\% of ECLAIRs GRBs. \\ 
& Fully covers ECLAIRs error boxes. \\ 
\botrule
\end{tabular}}
\label{tab:vt}
\end{table}

\subsection{Ground-based Instruments}
\label{sub:ground}

\subsubsection{Ground Wide Angle Cameras -- GWAC}
\label{ssub:gwac}

\begin{figure}[ht]
\begin{center}
\includegraphics[width=5.in]{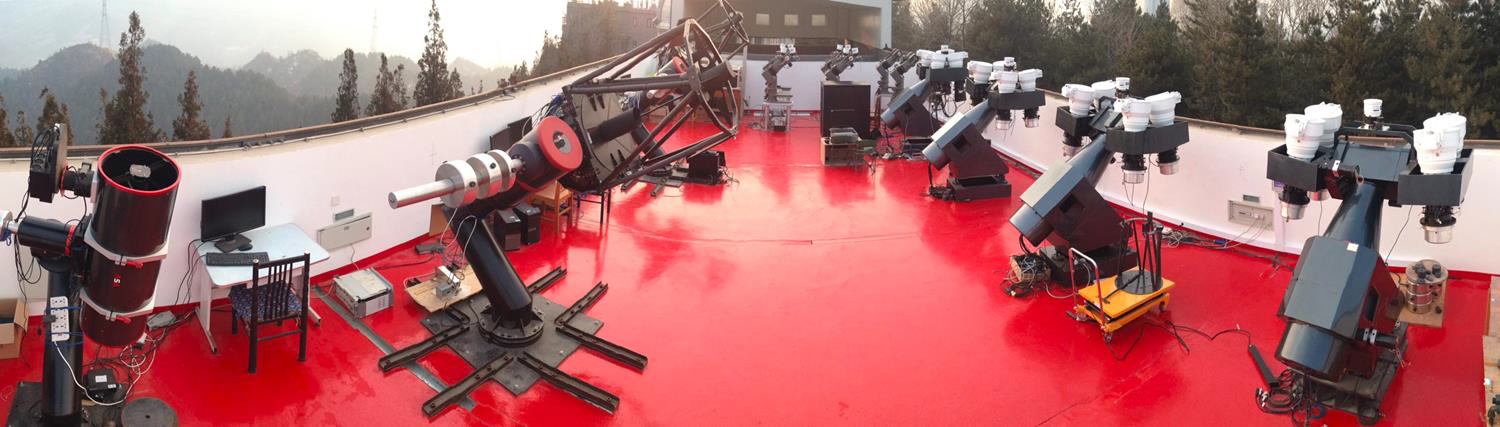}
\caption{Four GWAC units at the Xinglong Observatory (on the right of the picture).}%
\label{fig:gwac}
\end{center}
\end{figure}

The Ground Wide Angle Cameras (GWAC, \fref{fig:gwac})\cite{Han2021} is a set of several units, each made of 4 wide-angle optical cameras (called JFoV) and a small photographic camera (called FFoV) on a single fast-moving mount (\fref{fig:gwac}). Each unit covers instantaneously about 12\% of the ECLAIRs field of view, five of them have been installed at Xinglong Observatory (China) and 4 more will be installed at Muztagh Ata Observatory (China). The characteristics of JFoV and FFoV cameras are given in \tref{tab:gwac}, the FFoV extends the optical flux coverage by $\sim6$ mag in R-band, at the bright end. GWACs will scan the entire accessible sky each night to detect optical transients, offering the unique capability to detect precursor optical emission for \textit{SVOM} GRBs. Three robotic telescopes on the same site (two 60~cm diameter and one 30~cm diameter) can quickly follow interesting events detected by GWAC.
The GWACs are in commissioning phase, they have already participated in the search for counterparts of gravitational wave events during the O2 and O3 runs of LIGO/VIRGO\cite{Turpin2020}.

\begin{table}
\tbl{GWAC main characteristics.}
{\begin{tabular}{ll}
\toprule
Number of units & 5 (Xinglong) + 4 (Muztagh Ata) \\
Number of telescopes per unit & 4 (JFoV) + 1 (FFoV) \\
Diameter (mm)  & 180 (JFoV) -- 35 (FFoV) \\
Detector  & 4k x 4k CCD (JFoV) -- 3k x 3k CCD (FFoV) \\
Field of View (degrees) & 12.8 x 12.8 (JFoV) -- 30 x 30 (FFoV)  \\
Sensitivity ($\sigma$) & R $\sim 16$ (JFoV) -- R $\sim 12$ (FFoV)  \\
Location & Xinglong and Muztagh Ata Observatories (China)  \\
Additional features & Self triggering capability. \\ 
& The FFoV performs guiding. \\
\botrule
\end{tabular}}
\label{tab:gwac}
\end{table}

\subsubsection{Ground Followup Telescopes -- GFT}
\label{ssub:gft}

\begin{figure}[ht]
\begin{center}
  \parbox{2.4in}{\includegraphics[width=2.3in]{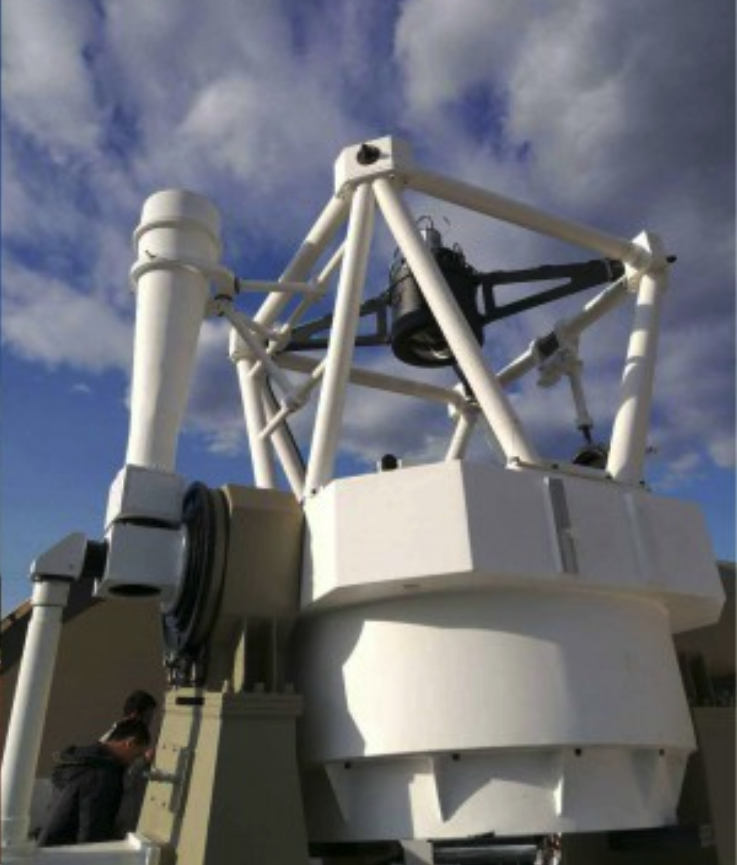}\figsubcap{a}}
  \hspace*{4pt}
  \parbox{2.4in}{\includegraphics[width=2.0in]{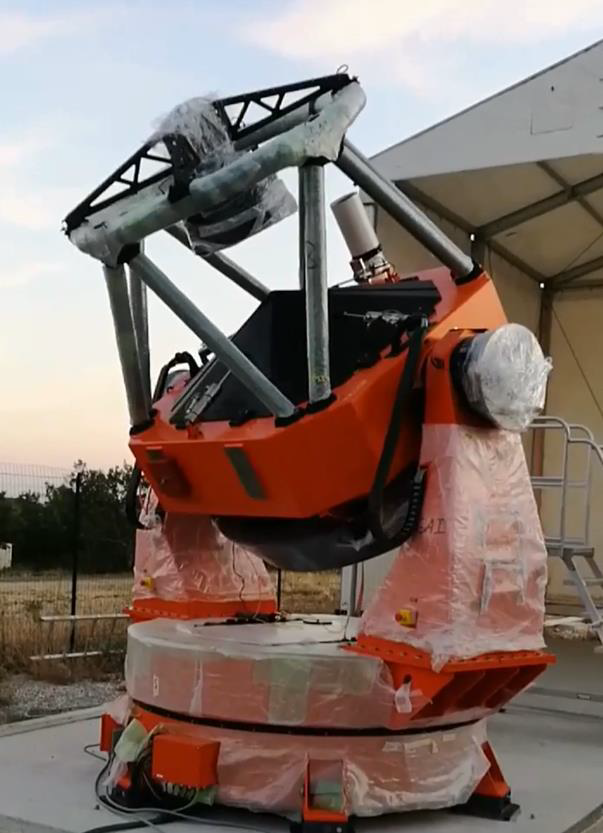}\figsubcap{b}}
  \caption{The two GFTs. (a) Chinese GFT at Jilin observatory. (b) French-Mexican GFT (aka Colibri) under test at the ``Observatoire de Haute Provence'' in France, before its installation at the Mexican National Astronomical Observatory in San Pedro M\'artir.}%
  \label{fig:gft}
\end{center}
\end{figure}

Two Ground Followup Telescopes (GFT, \fref{fig:gft}) aim at observing quickly the high-energy transients detected by \textit{SVOM}, to precisely measure their light curves and their positions. The two GFTs are respectively located in China and Mexico, about 120\degree\ apart in longitude, to maximise the probability of immediate observations in response to \textit{SVOM} alerts. Upon reception of an alert, the GFT which is in the night (if any) will start observing the source - if it is above the horizon - in less than 1 minute. 
The GFTs will also be used to perform joint observations with the satellite as well as non-\textit{SVOM} observations.
The main characteristics of the Chinese GFT (CGFT\footnote{https://www.svom.eu/en/\#filter=.instrumentsarticle.portfolio217load}) and of the French-Mexican GFT (Colibri\cite{Corre2018}) are given in \tref{tab:gft}.

\begin{table}
\tbl{GFTs main characteristics.}
{\begin{tabular}{lcc}
\toprule
& C-GFT & Colibri \\
\colrule
Diameter (mm)  & 1200 & 1300 \\
Focal ratio & 8 & 3.75  \\
Number of channels & 3 (g ; r ; i) & 3 (g/r/i ; z/y ; J/H) \\
Field of View (arcminute) & 21 x 21 & 26 x 26 (grizy) ; 21 x 21 (JH) \\
Sensitivity (r channel, 300~s, 10$\sigma$) & m$_{AB} \approx 20$ & m$_{AB} \approx 22$  \\
\botrule
\end{tabular}}
\label{tab:gft}
\end{table}

\subsection{The Mission}
\label{sub:mission}

\subsubsection{Orbit and Pointing Strategy}
\label{ssub:pointing}
\begin{figure}[ht]
\begin{center}
  \parbox{2.8in}{\includegraphics[width=2.7in]{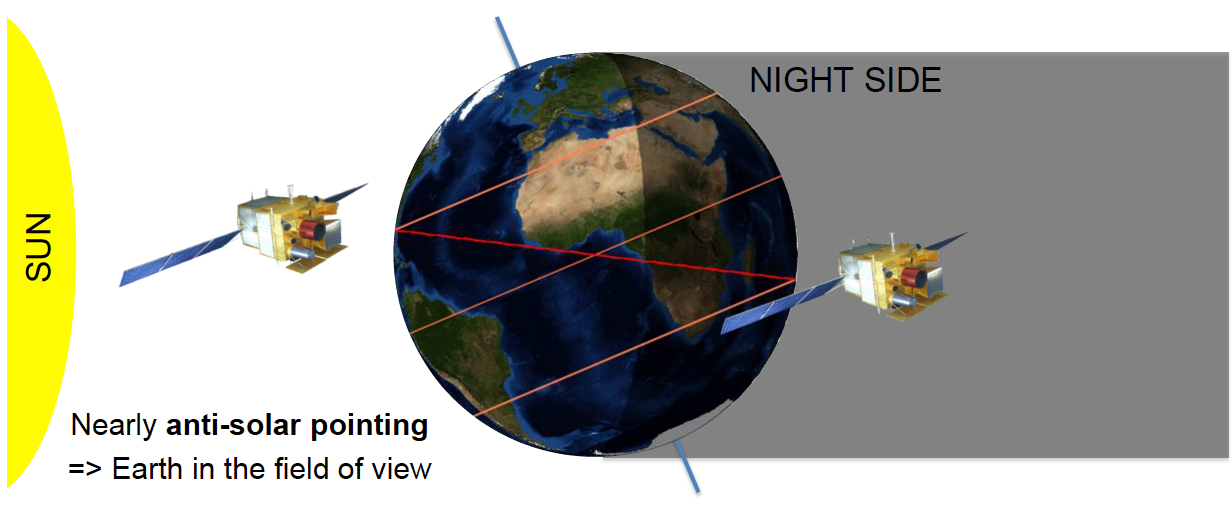}\figsubcap{a}}
  \hspace*{4pt}
  \parbox{2.in}{\includegraphics[width=1.9in]{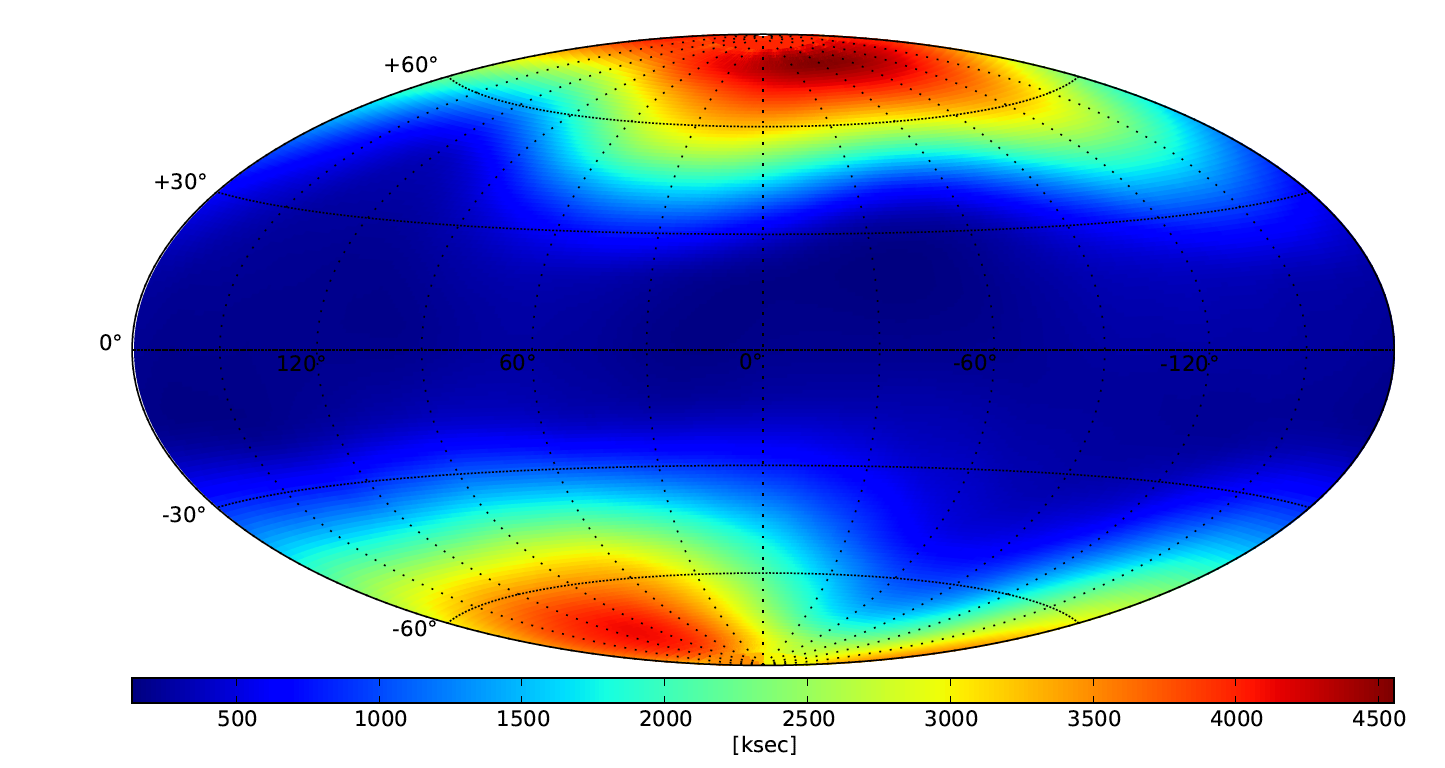}\figsubcap{b}}
  \caption{\textit{SVOM} pointing strategy. (a) Illustration of the anti-solar pointing that allows telescopes on the night side of the Earth to observe SVOM alerts promptly. (b) ECLAIRs one year exposure map in galactic coordinates.}%
  \label{fig:pointing}
\end{center}
\end{figure}

The \textit{SVOM} satellite will be launched by a Long March 2C launch vehicle, on a 29\degree\ inclined orbit, at an altitude of 630~km.
The pointing strategy has been designed to favour the detection of extragalactic transients and their immediate observability by ground based telescopes. This calls for a nearly anti-solar pointing (\fref{fig:pointing}a) with some excursions to avoid SCO-X1 and the galactic plane. In the following this pointing strategy is called ``the B1 law''. The resulting 1 year sky coverage of ECLAIRs is shown in \fref{fig:pointing}b, it favours the galactic poles, with exposure times reaching 3 to 4~Ms per year. 
The B1 law will be the rule during the nominal part of the mission (the first three years), with three exceptions: to point GRBs detected onboard, to point targets of opportunity, and to observe few remarkable sources as part of the general program (cf. \Sref{ssub:obsprog}).

\subsubsection{Observing Programs}
\label{ssub:obsprog}

\begin{figure}[ht]
\begin{center}
\includegraphics[width=5.in]{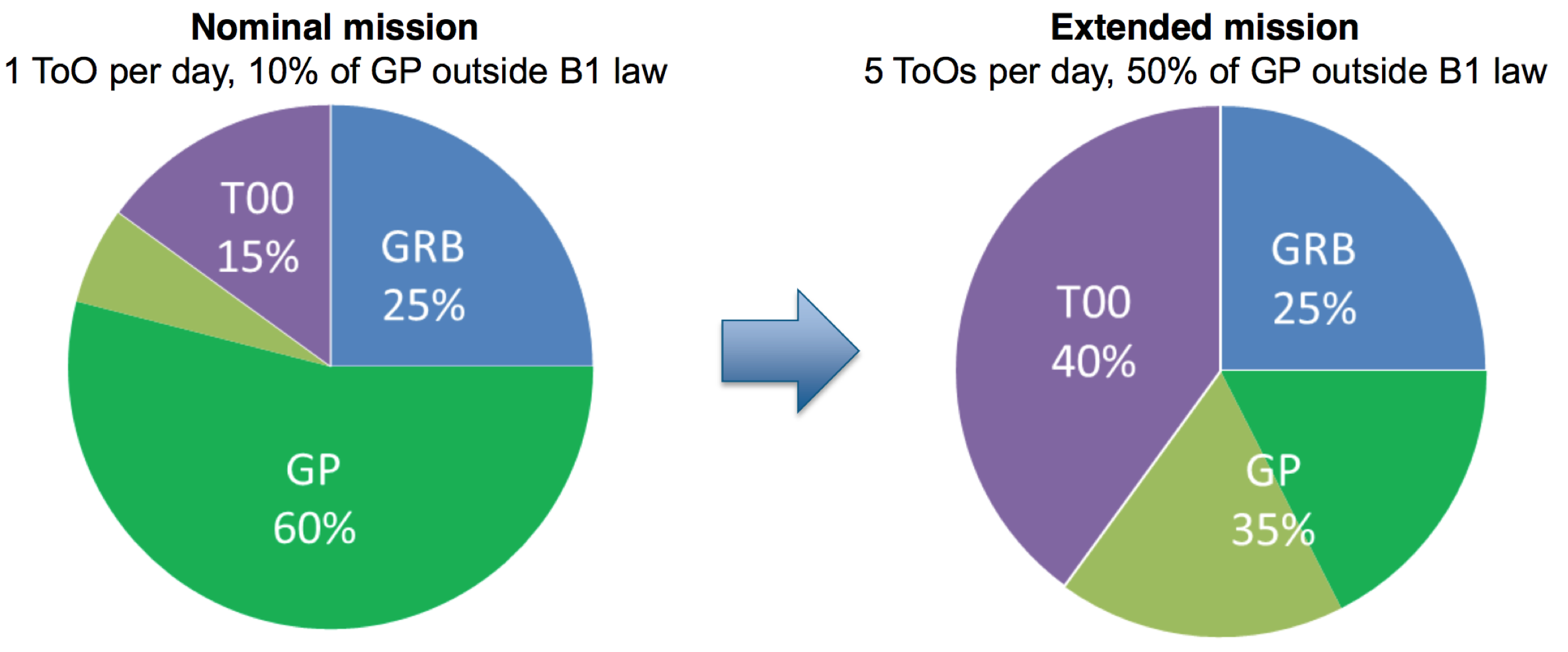}
\caption{Fraction of observing time attributed to the three observing programs of SVOM. Left: during the nominal 3-yr observing phase. Right: during the 2-yr extended phase. Light green indicates the time devoted to the General Program outside the B1 law.}%
\label{fig:obsprog}
\end{center}
\end{figure}

The observing time of \textit{SVOM} is divided into three components: the \textit{Core Program} (CP), which is the observation of HE transients detected onboard, the \textit{General Program} (GP), which is the observation of pre-selected sources close to the B1 law with the narrow field instruments, and the \textit{Target of Opportunity} program (ToO) to observe active sources upon request from the ground\cite{Wei2016}. The GP is selected by a TAC every six months and uploaded every week for the next two weeks. The ToOs are approved by the PIs of the mission and uploaded as soon as possible (see \sref{ssub:alert}). \Fref{fig:obsprog} shows the time allocated to each of the three programs during the nominal mission and during the extended mission.

\subsubsection{Alerts and ToOs}
\label{ssub:alert}
\begin{figure}[ht]
\begin{center}
\includegraphics[width=5.in]{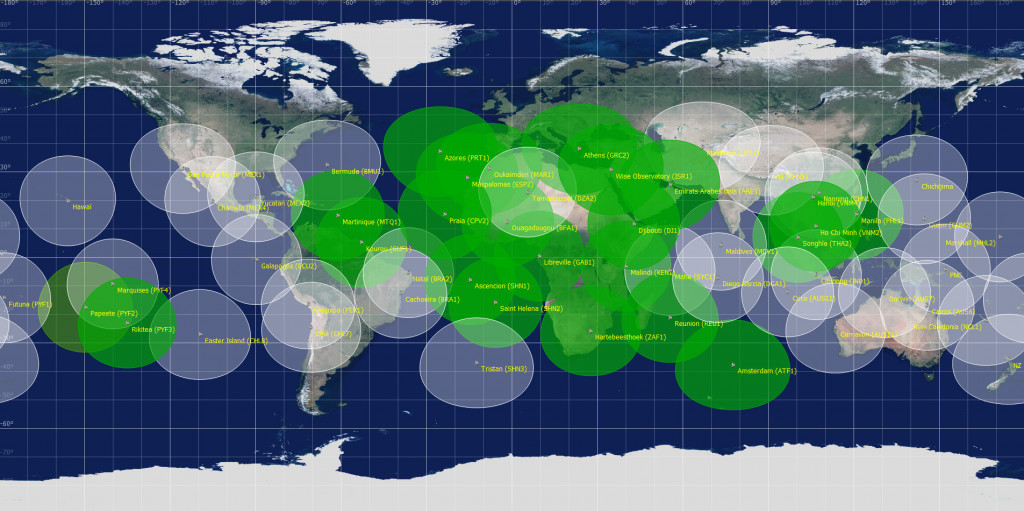}
\caption{\textit{SVOM} network of VHF receiving stations. In green the stations installed and fully operational at the end of 2021.}%
\label{fig:alert}
\end{center}
\end{figure}

\textit{SVOM} alerts will be distributed promptly to the community through the GCN network. Alerts can report a count rate excess or the detection of a new source or both. The detection of a count rate excess triggers the construction of a sky image and the search for a new source. A catalog of known sources onboard permits differentiating new transients from known sources\cite{Dagoneau2021}, the slew threshold for known sources is set very high in order to slew towards a known source only when it is in a truly exceptional state. Alerts validated onboard will be broadcast by two onboard VHF antennas, hopefully detected by one of the $\sim 45$ VHF receiving stations distributed around the world (\fref{fig:alert}) and forwarded to the French Science Center, which will format them as GCN Notices and VOEvents and transmit them immediately to the GCN\cite{Barthelmy1998}.

On the way up, commands can be sent quickly to \textit{SVOM} with the Beidou short messages communication service\cite{Beidou2021}. This will be used to program target of opportunity observations towards active high-energy sources or towards neutrinos or gravitational waves transient sources.

\subsubsection{Burst Advocates and Instrument Specialists}
\label{ssub:isba}

Like other GRB missions (e.g. \textit{Swift} or \textit{Fermi} ), the management of \textit{SVOM} alerts will be assigned to \textit{Burst Advocates} who will survey all activities connected with the alert, act as points of contact for the non-\textit{SVOM} world, and eventually take decisions about the follow-up of the source.
Considering the large number of space and ground instruments involved in the detection and follow-up of transients and the number of burst advocates needed to ensure the proper follow-up of the alerts, they cannot be specialists of all the instruments of SVOM. It is thus needed to complement the pool of burst advocates with a few \textit{Instrument Specialists} (2 or 3 per instrument) that can be contacted at any time by a burst advocate who has questions about the details of an instrument's operation.

\subsubsection{Mission Status}
\label{ssub:status}

All \textit{SVOM} instruments have been constructed, they have undergone environment testing and full characterization, which have demonstrated their nominal performance. They are ready to be shipped to Shanghai to be integrated on the satellite in the first half of 2022. The full satellite will then undergo complete testing, before the launch planned mid 2023.
The complex \textit{SVOM} system (satellite operations and programming, observing sequences with and without triggers, satellite slews, alert reception and distribution, data management, etc.) has also been tested during extensive ``system tests'' lasting several days, that reproduced realistic operating conditions. These tests have demonstrated the capability of the \textit{SVOM} collaboration to manage all aspects of this complex mission.

\section{\textit{SVOM} \& GRB Science}
\label{sec:svsci}

We conclude this presentation with a subjective list of GRB science topics that could make good progress in the coming years thanks to \textit{SVOM}. 
A presentation of the complete list of scientific objectives of the mission can be found in Wei et al.\cite{Wei2016}


\subsection{GRB detection}
\label{sub:detection}
Thanks to its low energy threshold and a thorough onboard image processing, ECLAIRs will be particularly sensitive to long and soft GRBs. 
As such, it provides a unique opportunity to increase the sample of X-Ray Flashes\cite{Barraud2003, Sakamoto2005} (XRFs) that may hold the key to the connection between SNIbc supernovae and classical long GRBs. 
This is also the guarantee to detect several high-redshift GRBs (z $\ge 5$) during the nominal mission\cite{Wei2016}, despite the relatively small size of ECLAIRs.
This is important because SVOM has the potential to quickly identify high-z GRBs as explained below.

The pointing strategy of the satellite will guarantee long periods of stable pointing, allowing the detection of long, faint transients like the ultra-long GRBs. A study by Dagoneau et al.\cite{Dagoneau2020} shows that \textit{SVOM} may double the number of detected UL-GRBs (\Fref{fig:mgbnd}b), allowing detailed multi-WL studies of these mysterious events and their host galaxies. 

Finally, all ECLAIRs and GRM photons are sent to the ground, allowing the construction of various delayed off-line triggers that will benefit of more computation power than available onboard and a much better knowledge of the context (solar activity, particles, pre- and post- count-rate evolution...), and could be used to search for classical HE transients with better sensitivity as well as new types of HE transients.

\subsection{Physics of the Prompt Emission}
\label{sub:physics}

\begin{figure}[ht]
\begin{center}
  \parbox{2.5in}{\includegraphics[width=2.4in]{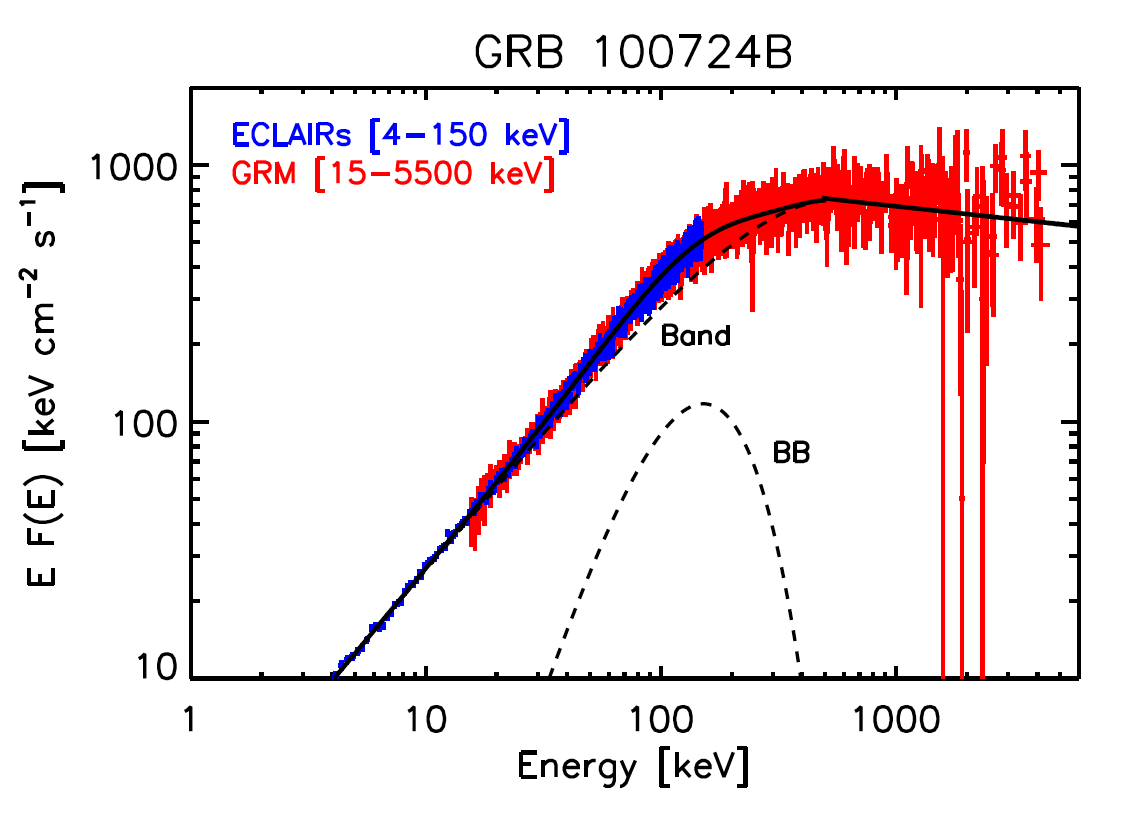}\figsubcap{a}}
  \hspace*{4pt}
  \parbox{2.2in}{\includegraphics[width=2.1in]{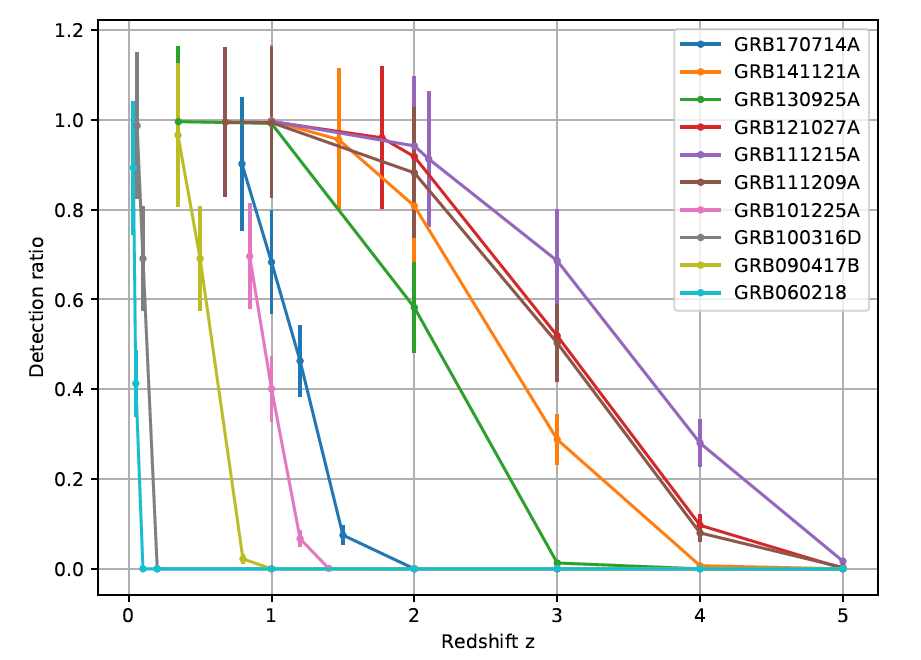}\figsubcap{b}}
  \caption{Predicted \textit{SVOM} performance. (a) Joint reconstruction of GRB spectra with ECLAIRs and GRM\cite{Bernardini2017}. (b) ECLAIRs sensitivity to ultra-long GRBs\cite{Dagoneau2020}.}%
  \label{fig:mgbnd}
\end{center}
\end{figure}

The unique combination of observations with GRM, ECLAIRs and GWAC for some GRBs will provide a unique coverage of the prompt emission. 
The systematic broadband time-resolved spectroscopy of the prompt GRB emission with ECLAIRs+GRM (and occasionally GWAC) will offer a renewed view of this phase, allowing to disentangle the role of the photosphere from internal shocks and other dissipation processes\cite{Bernardini2017} (\Fref{fig:mgbnd}a). 
For some GRBs, the prompt coverage may be extended to VHE gamma-rays with Fermi/LAT and CTA, allowing crucial diagnostic on the physics of GRB relativistic jets\cite{magic2019, hess2019}. 

\subsection{Physics of the Afterglow}
\label{sub:afterglow}

\textit{SVOM} has two features that will permit to study the physics of GRB afterglows systematically with more detailed data. The first one is a good coverage of the prompt-afterglow transition for GRBs longer than 2-3 minutes, in X-rays thanks to the overlap of ECLAIRs and MXT energy ranges, and in the optical thanks to the combination of GWAC + GFT + VT. The second is the multi-wavelength coverage of the afterglow with MXT, VT and GFT, with a good sensitivity balance between VT \& MXT, ensuring that $\sim70\%$ of the afterglows of \textit{SVOM} GRBs will be detected by both instruments.
In addition, it is expected that most \textit{SVOM} GRBs will have their redshift measured (see below), giving us access to the energetics of the bursts.

\subsection{Redshifts of GRBs}
\label{sub:sources}

\textit{SVOM} will provide arcsecond GRB localizations in less than 5 minutes with the GFTs and the VT, and the majority of \textit{SVOM} GRBs will be immediately observable with telescopes in the night hemisphere. These localizations will permit fast spectroscopic observations with mid- or large-size telescopes with the goal of measuring the redshifts of a large fraction ($\geq 2/3$) of \textit{SVOM} GRBs.
 \textit{SVOM} will also provide quick indications of intermediate redshift GRBs\cite{Wang2020} ($3 \leq \mathrm{z} \leq 6$), and of high-z GRBs candidates, when the afterglow is undetected in the VT. In all cases, dark bursts in the VT will indicate GRBs of interest, deserving additional observations with powerful telescopes.

\subsection{SN-less GRBs}
\label{sub:snless}
\begin{figure}[ht]%
\begin{center}
\includegraphics[width=3.5in]{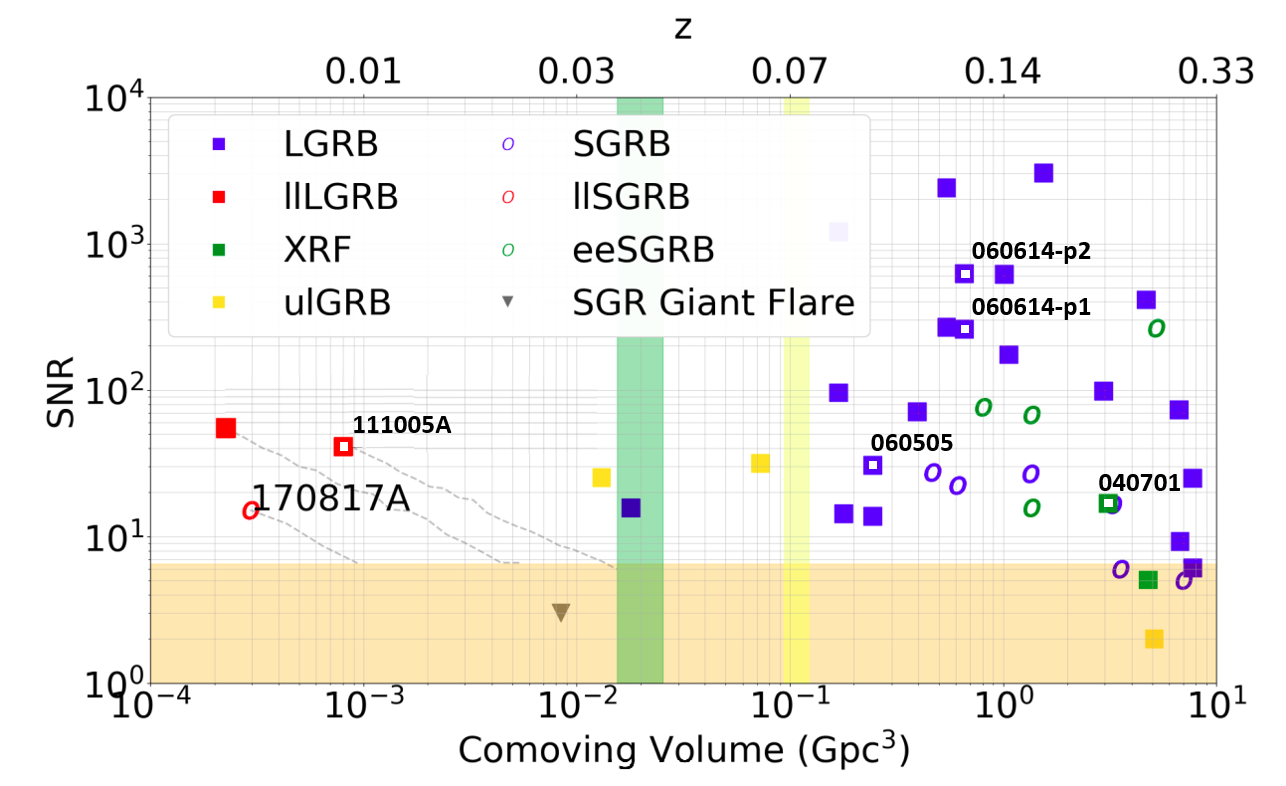}
\caption{ECLAIRs on-axis count SNR for GRBs in the local universe (z$\leq 0.3$). The orange horizontal band represents the detection limit of ECLAIRs at SNR = 6.5. The green and yellow bands represent the O4 LIGO detection limits for NS-NS and BH-NS mergers respectively. The light grey trails represent the evolution of the on-axis count SNR with the redshift\cite{Arcier2020}.}
\label{fig:ba20}
\end{center}
\end{figure}

The detection of two SN-less GRBs with \textit{Swift} in 2006\cite{DellaValle2006,Fynbo2006,GalYam2006,Gehrels2006} raised a lot of attention, since the association of long GRBs with SNIbc was by then considered secure. In 2011, another SN-less GRB was detected by \textit{Swift}: GRB~111005A\cite{Michalowski2018} at 57 Mpc (z=0.01326). These events raise the question of the nature of SN-less GRBs: are they the result of the explosion of the core of a massive star or a merger?
Since GRB~111005A was well within the detection volume of GW detectors for NSNS or BHNS mergers and easily detectable with ECLAIRs (\Fref{fig:ba20}), the occurrence of another event of this type within the field of view of ECLAIRs during the operation of GW interferometers would provide crucial data to solve this mystery\cite{Arcier2020}.

\section{Conclusion}
\label{sec:conclusion}

The operation of \textit{SVOM} simultaneously with powerful instruments in the 2020's decade will certainly bring new crucial data for our understanding of GRBs. We have already mentioned the diagnostic brought by gravitational waves for nearby events, like short GRBs, sub-luminous GRBs or XRFs. We can also mention the search for GRBs coincident in time and direction with afterglow candidates that will be detected by the Vera Rubin Observatory. The operation of \textit{SVOM} simultaneously with large neutrino detectors like ICECUBE or KM3NeT may also bring its share of surprises. Last but not least, the discovery of VHE gamma-ray emission from GRBs by HESS and MAGIC raises the hope to increase the number of such detections with CTA in the SVOM era, revealing key information about the composition and physical processes at work in ultrarelativistic jets.

Observing GRBs, AGNs, TDEs and transient sources of gravitational waves, SVOM will become a key player in the fields of High-Energy Astrophysics, Time Domain Astronomy and Multi-Messenger Astrophysics, after its launch mid-2023.

\bibliography{SVOM_MG16}

\end{document}